# Knowledge Engineering for Open Science: Building and Deploying Knowledge Bases for Metadata Standards

Mark A Musen, Martin J. O'Connor, Josef Hardi, Marcos Martínez-Romero
Stanford Center for Biomedical Informatics Research
Stanford University School of Medicine
Stanford, CA  94305
*musen@stanford.edu*

## Abstract

Background: For more than a decade, scientists have been striving to make their datasets available in open repositories, with the goal that they be findable, accessible, interoperable, and reusable (FAIR).  Although it is hard for most investigators to remember all the "guiding principles" associated with FAIR data, there is one overarching requirement: The data need to be annotated with "rich," discipline-specific, standardized metadata that can enable third parties to understand who performed the experiment, who or what the subjects were, what the experimental conditions were, and what the results appear to show.  Most areas of science lack standards for such metadata and, when such standards exist, it can be difficult for investigators or data curators to apply them.

Methods: The Center for Expanded Data Annotation and Retrieval (CEDAR) builds technology that enables scientists to encode descriptive metadata standards as *templates* that enumerate the attributes of different kinds of experiments and that link those attributes to ontologies or value sets that may supply controlled values for those attributes. These metadata templates capture the preferences of groups of investigators regarding how their data should be described and what a third party needs to know to make sense of their datasets.

Results:  CEDAR templates describing community metadata preferences have been used to standardize metadata for a variety of scientific consortia.  They have been used as the basis for data-annotation systems that acquire metadata through Web forms or through spreadsheets, and they can help correct metadata to ensure adherence to standards.

Conclusion:  Like the declarative knowledge bases that underpinned intelligent systems decades ago, CEDAR templates capture the knowledge of a community of practice in symbolic form, and they allow that knowledge to be applied in a variety of settings.  They provide a mechanism for scientific communities to create shared metadata standards and to encode their preferences for the application of those standards, and for deploying those standards in a range of intelligent systems to promote open science.

**Keywords**: open science, standards, FAIR guiding principles, knowledge engineering, metadata, CEDAR Workbench

## Introduction

Feigenbaum (1984) famously referred to the construction of knowledge bases for intelligent systems as "the applied side of artificial intelligence."  More than four decades later, that view seems rather restricted, as AI has moved on in many ways since the era of large, symbolic knowledge bases that are meticulously created by hand.   Nevertheless, declarative knowledge representations that are inspectable and editable by humans and that power intelligent computer systems retain an important place in the world, and they are essential in application areas where there are few data from which to drive machine learning.  Such knowledge bases are also essential for capturing nuanced distinctions and for serving as a blackboard through which a professional community can work to achieve consensus on how people—and computers—should act in particular situations.

In this paper, we show how such knowledge bases can be extremely valuable for encoding and communicating how groups of scientists believe research data should be shared.  Although the idea of sharing scientific data may seem a bit arcane, the ability to access research data openly and widely is increasingly seen as essential to the scientific enterprise.   There is growing expectation that third parties should be able to retrieve the online research results of other investigators to verify experimental claims and to make new discoveries.  After the publication of a landmark paper by Wilkinson and his colleagues (2016), it has become widely accepted that research datasets should be findable, accessible, interoperable, and reusable (FAIR).  Scientists are abuzz with the idea that their datasets should be FAIR—both out of a conviction that FAIR data are important for the research enterprise and because their publishers and sponsors require it (Tollefsin and Van Noorden, 2022).

Wilkinson et al. (2016) enumerated a collection of "guiding principles" needed for datasets to be FAIR.  Although the FAIR principles can seem obscure, they are dominated by a single, rather simple idea:  FAIR data need to be annotated with descriptive metadata that are "rich" and that adhere to relevant community standards (Musen et al., 2022).  The metadata need to provide information about the scientific context of the work (e.g., the investigators, the subjects, the experimental conditions, and the reported results) and this information needs to be understandable both to humans and to machines.

We can view descriptive metadata as a comprehensive list of attributes of the experiment (tailored to the research domain), including the interventions that are performed and the kinds of observations that are made.  These attributes, taken together, comprise a *reporting guideline*—a standard set of descriptors for a given class of experiment.  The FAIRSharing resource (Sansone et al., 2019) provides details for more than 300 such reporting guidelines for standardizing metadata across many branches of science.  When investigators create a metadata record, they assign a value to each attribute in the relevant reporting guideline.  They thus may say, informally, that the **subject** of the experiment was a *mouse*, that the experiment involved an examination of the **tissue** in the mouse's *liver*, that the liver was examined after an **intervention** that consisted of the administration of some *drug*, and so on.  Thus, the metadata can be viewed as a list of attribute–value pairs.  For a dataset to be FAIR, at minimum it must include metadata that adhere to a standard reporting guideline (i.e., the metadata must include the correct set of attributes for describing the experiment) and each metadata attribute must have a corresponding value that adheres to the correct datatype (i.e., when appropriate, a term from a standard value set or ontology).  When the attributes and values in research metadata adhere to such standards, the datasets are more FAIR because searches for metadata have greater recall and higher precision, and the datasets are more interoperable and reusable because there is (hopefully) no guesswork regarding what experiment was done and what the corresponding data mean.

But there's a problem.  Despite the advantages of having standardized metadata for describing datasets, there are relatively few areas of science where such standards exist, considering the many areas of investigation and numerous kinds of experiments that researchers may perform.  Moreover, even when there are available standards, investigators are often terrible at ensuring that their metadata adhere to those standards (Gonçalves and Musen, 2019).  Most online experimental results are annotated in idiosyncratic, unsearchable ways, making it virtually impossible for third parties to gain any value from the data (Musen, 2022).  There is a groundswell of desire to make scientific datasets "AI ready," but that will never happen until scientific datasets are made FAIR through adherence to metadata standards, and that will never happen until such standards are more universal and until investigators have the ability to apply such standards in a more straightforward manner.

Our team at Stanford University has been working for the past decade to create the infrastructure needed to support the development and application of machine-actionable standards for scientific metadata.  At the core of our approach are declarative knowledge bases that correspond to templates for specifying discipline-specific metadata standards in a structured, consistent manner.  Our work points the way for a generalizable infrastructure that can ensure data FAIRness, and it demonstrates that the lessons that the AI community learned from knowledge-engineering activities decades ago are still highly relevant in the era of open science.

## Materials and Methods

The Center for Expanded Data Annotation and Retrieval (CEDAR) was established in 2014 as part of the NIH Big Data to Knowledge initiative (Musen et al., 2015).  At the core of our activity has been the CEDAR Workbench (often simply referred to as "CEDAR"), which offers an integrated suite of software tools that allow developers to create metadata templates that reflect reporting guidelines for scientific experiments and that allow investigators and data curators to easily create standards-adherent metadata that comport with those guidelines (O'Connor et al., 2016).  A dedicated template-editing environment allows users to enumerate the attributes that should be specified in different metadata reporting guidelines, creating templates for the subsequent entry of dataset-specific metadata.  In the CEDAR Workbench, the system uses metadata templates to generate on the fly Web forms that scientists can use to enter instances of metadata (Figure 1).  The system restricts entries in each field of the form to the appropriate data type.  For fields whose values are to be taken from pre-enumerated value sets or ontologies, CEDAR automatically constructs a drop-down menu to enable selection of the applicable controlled term.  Thus, the CEDAR Workbench facilitates the construction of machine-actionable, human-understandable, standards-based metadata specifications and enables investigators to use those specifications to enter experiment-specific metadata that adhere to those standards in a straightforward manner.

The ontologies and value sets that CEDAR uses come from BioPortal, an open repository of nearly all the world's publicly available biomedical ontologies and controlled terminologies (Vendetti et al., 2025).  Although BioPortal's contents, by design, are biomedical, other investigators are building ontology repositories designed for other areas of science that adopt BioPortal's code base and APIs (Jonquet et al., 2023).  CEDAR not only presents lists of controlled terms from BioPortal, but also it can present controlled terms from standard online naming authorities (O'Connor et al., 2025a), such as ORCID (for the names of researchers), ROR (for the names of research organizations and institutions), and RRID (for the names of research resources, including reagents, chemicals, and strains of laboratory animals).

CEDAR uses a *template model* for encoding the reporting guidelines that scientific communities develop to ensure that the metadata that investigators use to annotate datasets are internally consistent, adherent to standards, and sufficient for third parties to make sense of the underlying data and to understand the nature of the experiment that was performed (Figure 2).  The standards reflected in CEDAR templates may not result from the arduous, systematized processes used by international standards development organizations such as ISO or ANSI, but they nevertheless reflect the explicit preferences of communities of investigators for how their data should be annotated. These *community* standards are what allow the corresponding community to search for research datasets online and know what was done to generate the data. They are what, at minimum, make the corresponding datasets FAIR.

**Figure 1:** Web form for creating metadata instances. The fields in the Web form are generated dynamically from a designated metadata template, in this case for a biological assay known as RNAseq. The user has provided values for the first four fields in the forms. The template indicates that values for the field "Analyte class" come from a predefined value set stored in the BioPortal repository, and the Web form thus displays a drop-down menu of possible selections. Entries in the Web form result in attribute–value pairs that are converted internally to JSON-LD, as shown in Figure 3.

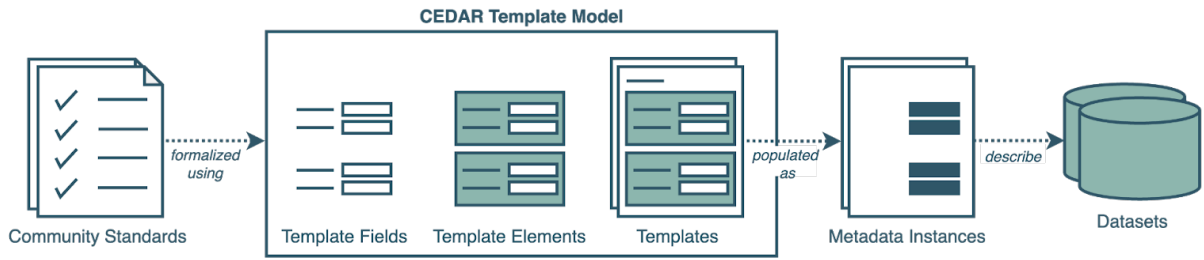

**Figure 2:** Relationships between the CEDAR template model, metadata templates, and metadata instances. Each CEDAR template is an instantiation of the template model and denotes a reporting guideline for some class of experiment. A CEDAR template consists of a set of fields (attributes) that may be grouped into template elements. Each metadata instance is a set of attribute–value pairs that represents an instantiation of some CEDAR template. Metadata annotate specific datasets, providing discipline-specific information about the experiment that led to the data.

At the core of the system is thus our underlying model for the specification of metadata templates. The CEDAR template model provides a framework for representing the structure, semantics, and constraints associated with scientific metadata. The model treats metadata standards that adhere to the model as first-class knowledge structures that can be authored, versioned, validated, and instantiated. In the CEDAR model, a *template* formally defines the attributes that constitute a reporting guideline for a particular kind of experiment or dataset, including the attribute names, data types, cardinalities, and, if indicated, links to the ontologies and value sets that may supply values for the attributes. A *metadata instance*, in turn, is a populated version of such a template—an annotation of a specific dataset with structured, standards-adherent information.

The core entities in a metadata template are *template fields* and *template elements*. Each template field corresponds to an attribute of a reporting guideline, and has a declared datatype (e.g., string, integer, date, controlled term, ROR, ORCID). Controlled-term fields can specify entire ontologies that might supply their values (e.g., Uberon, ChEBI, NCBITaxon), specific branches of an ontology (e.g., the "cell type" subtree of the Cell Ontology), curated value sets (e.g., analyte class; see Figure 1), or selected ontology terms (e.g., the classes "Male", "Female", and "Unknown" from NCBITaxon). All fields are encoded with human-readable labels and machine-resolvable IRIs, enabling both validation and semantic traceability. Fields can be grouped into reusable template *elements*, and multiple elements can be composed into a template. Each of these components is itself an addressable, reusable object in the CEDAR ecosystem, with a unique identifier and metadata about authorship, provenance, and versioning. Because fields and elements are first-class, reusable entities in CEDAR, commonly used template components (e.g., contributor information, sample descriptors, tissue provenance information) can be authored once and reused across multiple templates, promoting uniformity in metadata modeling.  In addition, the CEDAR template model allows for rich annotations of fields and templates—including descriptions, units, expected formats, and prompts and hints that can be displayed by the system when templates are instantiated—guiding users during metadata entry and ensuring consistency across metadata records.

CEDAR represents metadata templates in JSON Schema and it encodes metadata instances (populated templates) in JSON-LD (Figure 3), enabling CEDAR templates to be easily converted to other widely used formats such as RDF and LinkML.  Each field value in an instance retains its connection to the corresponding field in the original template, including any semantic constraints or references to ontologies from which the value is taken.  Thus, for fields whose values come from external ontologies or value sets, the instance includes both the human-readable label and the IRIs of the controlled terms from the reference vocabularies.  This dual encoding ensures that the metadata are not only interpretable by humans but also fully resolvable and interoperable in Semantic Web contexts.  The result is a richly annotated metadata record that can be programmatically validated, integrated with other linked datasets on the Web, indexed for discovery, and used as input for analytic workflows.  Because each instance is grounded in a well-defined template with explicit semantics, it becomes possible to align, compare, and federate metadata across repositories and domains with minimal ambiguity.  The representational choice makes it easy to integrate CEDAR templates with knowledge graphs and collections of linked data.

The scientists and data curators who populate CEDAR metadata templates are completely insulated from the underlying representation.  Domain specialists see the template only as rendered dynamically by the metadata authoring system (e.g., as in Figure 1), and they never need to worry about how metadata entries are encoded.  The metadata templates in CEDAR thus can be authored once, often in a collaborative fashion, using simple Web-based tools where representational details are exposed, and then the templates can reused by any number of researchers to capture actual metadata via tools that shield the scientists from the arcane aspects of knowledge representation and that create the necessary metadata-entry forms as needed directly from the templates.

## Results

CEDAR templates encoded using the CEDAR template model have been used in a variety of applications for several years.  We now present some of those applications, demonstrating

```json
{
  "schema:name": "RNAseq HBM296.DXLM.434 metadata",
  "@id": "https://repo.metadatacenter.org/template-instances/2801b14e",
  "schema:isBasedOn": "https://repo.metadatacenter.org/templates/944e5fa0",
  "@context": {
    "rdfs"   : "http://www.w3.org/2000/01/rdf-schema#",
    "xsd"    : "http://www.w3.org/2001/XMLSchema#",
    "schema" : "http://schema.org/",
    "oslc"   : "http://open-services.net/ns/core#",
    "skos"   : "http://www.w3.org/2004/02/skos/core#",
    ...
  },
  "parent_sample_id": {"@value": "HBM296.DXLM.434"},
  "lab_id": {"@value": "3432_ftr_RNA_A2"},
  "preparation_protocol_doi": {
    "@id": "https://dx.doi.org/10.17504/protocols.io.4r3l224p3l1y/v1"
  },
  "dataset_type": {
    "@id"        : "https://purl.humanatlas.io/vocab/hravs#HRAVS_0000310",
    "rdfs:label" : "RNAseq"
  },
  "analyte_class": {
    "@id"        : "https://purl.humanatlas.io/vocab/hravs#HRAVS_0000327",
    "rdfs:label" : "DNA + RNA"
  },
  "acquisition_instrument_vendor": {
    "@id"        : "https://identifiers.org/RRID:SCR_023672",
    "rdfs:label" : "10x Genomics"
  },
  "acquisition_instrument_model": {
    "@id"        : "https://identifiers.org/RRID:SCR_017202",
    "rdfs:label" : "BZ-X710"
  },
  "source_storage_duration_value": {"@value": "10", "@type": "xsd:decimal"},
  "source_storage_duration_unit": {
    "@id"        : "http://purl.obolibrary.org/obo/UO_0000032",
    "rdfs:label" : "hour"
  },
  ...
}
```

**Figure 3:** JSON-LD representation of a metadata instance. This particular representation encodes a small portion of the metadata values for the instance depicted in Figure 1. The metadata describe the details of an RNAseq experiment, whose data are being annotated with these metadata. Note that `analyte_class` has a value from the list of controlled terms shown in Figure 1. `Acquition_instrument_vendor` and `acquisition_instrument_model` have values that are specific research resource identifiers (RRIDs).

the advantages of using a core model to represent metadata standards in a reusable, machine-actionable manner.  This diversity of implementations underscores how a reusable metadata model—by abstracting community standards into structured, machine actionable templates—can support consistent, scalable, and interoperable metadata practices across a broad range of scientific domains and technical environments.  Once we have settled on a reusable metadata model, that model can form the foundation for a wide range of practical applications that can have a dramatic effect on the management of scientific data in the real world.  More important, rendering a scientific community's metadata standards as a set of declarative knowledge bases makes those standards inspectable, shareable, editable, and reusable—thus streamlining the management and propagation of the groundwork that underlies FAIR data.

The CEDAR Workbench

The CEDAR Workbench is an integrated platform that includes components for (1) creating, editing, storing, and sharing metadata templates that comport with the CEDAR template model, (2) selecting a metadata template and instantiating it with standards-adherent values, (3) selecting a target repository and uploading the instantiated metadata along with the associated research data, thereby creating a dataset archive.  All components are implemented as a collection of microservices, each accessible via a HTTP-based API, and the full set of CEDAR features is accessible via a Web-based user interface.  The Workbench has been used by a variety of research consortia for standards-adherent metadata management.

A good example is the NIH initiative for Helping to End Addiction Long-term (HEAL), an enormous activity that is addressing the opioid crisis at all levels.  HEAL involves more than one thousand projects to identify new therapeutic targets for treating pain and substance use disorder, to develop nonpharmacological strategies for pain management, and to advance overdose and addiction treatment (Sawyer-Morris et al., 2025). An important goal of the HEAL initiative is for experimental data to be open and FAIR, enabling investigators to access and reuse data quickly in a rapidly accelerating research environment.

HEAL uses an extensive data-management infrastructure developed at the University of Chicago, along with the original CEDAR Workbench to ensure that datasets are annotated with adequate, discipline-specific metadata (Figure 4).  The consortium has worked completely independently to access the CEDAR Workbench, to create metadata templates appropriate for HEAL studies, and to provide extensive training materials to the HEAL investigators to help them to annotate experimental datasets in compliance with standards developed by the community. The HEAL investigators have become adept at metadata editing using CEDAR, and their use of shared metadata specifications reportedly has offered the scientists in the consortium considerable advantages in building off one another's work.

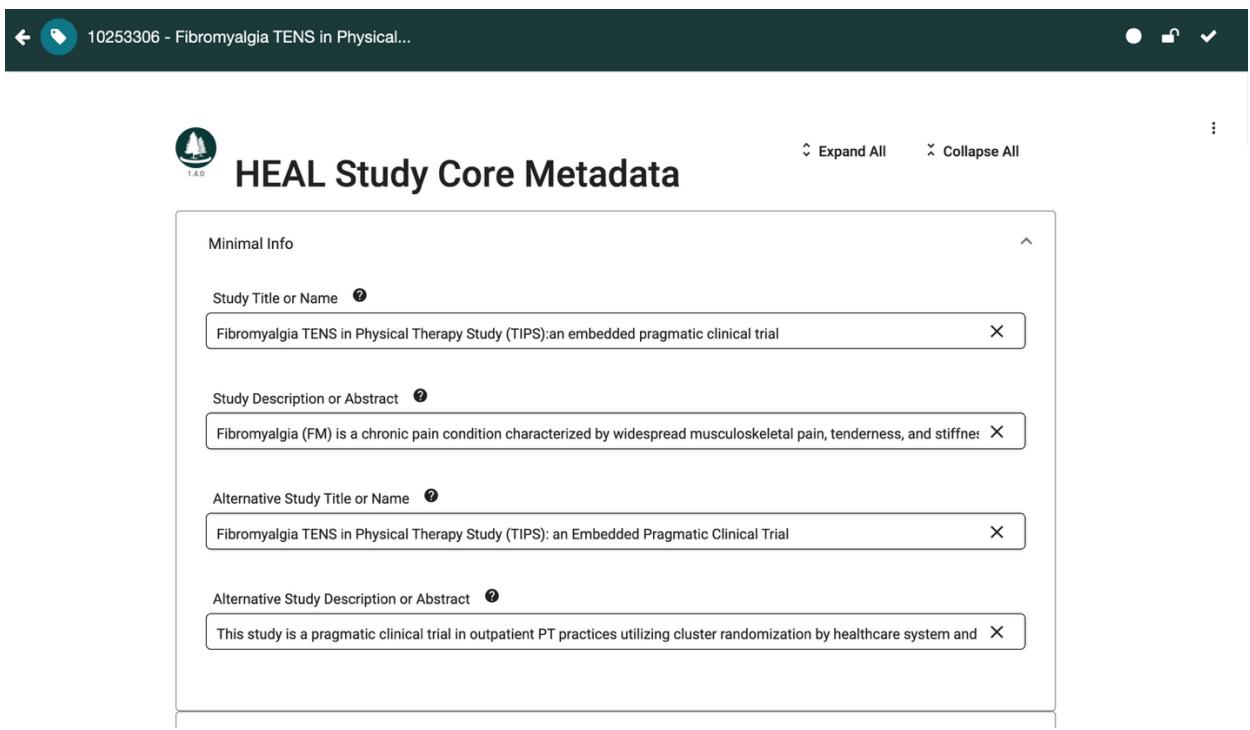

**Figure 4:** The CEDAR Workbench as used by the HEAL consortium. The figure shows CEDAR as used to acquire structured metadata describing a clinical trial, in this case a study of transcutaneous electrical nerve stimulation (TENS) for management of fibromyalgia.

Another NIH consortium, one that attempts to characterize the functions and properties of proteins and genes that may serve as good targets for new drugs ("Illuminating the Druggable Genome"; IDG), is also a user of the CEDAR Workbench (Vidović et al., 2024; Egyedi et al., 2017). IDG data curators use the CEDAR Workbench to validate and edit metadata records to ensure that dataset annotations comport with appropriate standards.

In the Netherlands, government agencies have taken decisive action to guarantee FAIR data. The Netherlands Agency for Health Research and Development (ZonMw), for example, often requires that the investigators whom it supports commit in advance to disseminating their research results openly in accordance with predetermined metadata standards *as a condition of funding* (Bloemers and Montesanti, 2020). A not-for-profit organization, known as Health-RI (2025), disseminates the investigators' datasets for use by the scientific community over the Web, making research results widely available and annotated through standardized metadata that researchers author using the CEDAR Workbench.

In all these systems, developers create metadata standards using CEDAR templates, and the CEDAR Workbench then guides end-users in instantiating the templates to create standards-adherent metadata and, by extension, FAIR datasets.  HEAL, IDG, and Health-RI templates all adhere to the CEDAR template model, and they thus can be used in any of the three systems.

Embedding CEDAR in Other Software

The CEDAR Workbench was developed as an integrated system.  Often, however, software engineers require more flexibility in how they include CEDAR functionality in an overall architecture.  We therefore have created the CEDAR Embeddable Editor as a lightweight, interoperable Web component that allows CEDAR metadata editing and display capabilities to be incorporated easily within third-party Web applications (O'Connor et al., 2025a).  With the CEDAR Embeddable Editor, developers can easily build CEDAR metadata-management functionality directly into their custom applications, reading CEDAR-compatible metadata templates and metadata instances from a library that the developers designate and allowing end users to display and edit CEDAR-compatible metadata from within the particular host system.  This approach has been particularly effective in allowing our collaborators to build CEDAR functionality into general-purpose data repositories that are not tied to any specific application area.

The Open Science Framework (OSF), for example, is an open-source, open-access platform that supports a community of over 870,000 investigators in all stages of managing the products of their research (Foster and Deardorff, 2017).  OSF is managed by the Center for Open Science, and it offers infrastructure for users to upload, archive, and disseminate their datasets in a transparent manner.  The OSF development team has integrated the CEDAR Embeddable Editor into their platform, presenting investigators with the ability to select conventional CEDAR metadata templates from a library and to author standards-adherent metadata as needed (Olsen and Corker, 2024).  The OSF implementation allows user to fill in CEDAR templates that contain *descriptive metadata* fields for recording information about their experimental setting and subjects as well as *structural metadata* fields that capture details about the configuration of the dataset itself (Figure 5).

**Figure 5:** OSF acquires standardized metadata using the CEDAR Embeddable Editor. The figure shows the template for acquiring Psych-DS metadata, a specification for sharing data in the social and behavioral sciences. Here the user is entering structural metadata about a dataset describing a psychological experiment involving visual memory and attention.

Similarly, Dryad, another very widely used general-purpose data repository, offers a library of metadata templates—all based on the CEDAR template model—and provides users with the CEDAR Embeddable Editor to fill in templates and to annotate their datasets in a standardized manner. Although the Dryad work was originally stimulated by a project to formalize metadata in cognitive neuroscience (Lippincott, 2024), the Dryad template library continues to grow in breadth. When users upload datasets to Dryad and enter keywords that describe their experiment, the system may automatically suggest a discipline-specific metadata template appropriate for the submission.

All OSF and Dryad metadata templates reflect the standard metadata model that governs the CEDAR Workbench. As with the templates used by HEAL and IDG and Health-RI, all OSF and Dryad metadata templates are fully interoperable with all other CEDAR-compatible data-management platforms.

Using CEDAR Templates to Create Custom-Tailored Spreadsheets

Although our systems can generate sophisticated Web forms directly from CEDAR metadata templates, many scientists despise the idea of filling out questionnaires online. The use of spreadsheets is more familiar and more natural to them, and they like that spreadsheets give them the option to enter metadata for several experiments at once. Working with a large NIH-supported consortium known as the Human Biomolecular Atlas Program (HuBMAP), which aims to create a master catalog of the biomarkers that distinguish every type of cell in the healthy human body (Jain et al., 2023), we have developed an approach that transforms CEDAR templates into intelligent spreadsheets, and that offers researchers a simple, well-established mechanism for entering their metadata (O'Connor et al., 2025b). CEDAR thus can use the same metadata templates that it uses to render Web forms to generate spreadsheets (Figure 6). Users select a metadata template from the CEDAR library, and the system generates a spreadsheet (either in Excel or as a Google sheet) in which each *column* corresponds to a separate metadata field from the template. To manage metadata from collections of multiple experiments, each *row* in the spreadsheet captures metadata entries from a different experiment of the same type. The datatype of each field type is stored in the CEDAR template, allowing the system to create a spreadsheet that, to the extent possible, enforces the correct datatype for each cell in the corresponding column.

The problem with spreadsheets is that users can cheat. They can easily override datatype restrictions, entering strings that are not part of predefined value sets or ontologies, or typing in floating-point numbers when the field value is supposed to be an integer.

**Figure 6:** A metadata-entry spreadsheet generated from a CEDAR template. The spreadsheet was created from the same template for RNAseq metadata used to create the Web form shown in Figure 1. Like the Web form, the spreadsheet uses a predefined value set to ensure that the selected value for "analyte class" adheres to the given standard.

Scientists, of course, enjoy this flexibility, but the creation of nonstandard metadata entries risks the generation of datasets that are not FAIR. When metadata do not adhere to standard reporting guidelines, datasets are not easily searchable, and interoperability of the metadata cannot be guaranteed. We therefore have created a *spreadsheet validator* that reviews the spreadsheet-based metadata, identifying possible errors and suggesting how the errors might be corrected (O'Connor et al., 2025b). The validator can perform these functions because it, too, has access to the CEDAR template that was used to create the spreadsheet in the first place, and the template provides the knowledge needed to analyze the spreadsheet entries and to determine potential modifications for any values that may deviate from the intended standard (Figure 7).

Summary

As use of the CEDAR software has become more widespread, the underlying technology has evolved, and the approach has taken on new capabilities. At its core, however, the CEDAR template model has not changed, allowing the generation of metadata descriptions that (1) adhere to community-based standards for their content and (2) are formatted in a standardized knowledge-representation system for their structure. Like cassettes that can be plugged into a variety of playback systems, CEDAR metadata templates are compatible with a range of software systems that each contribute to data management tasks in different ways.

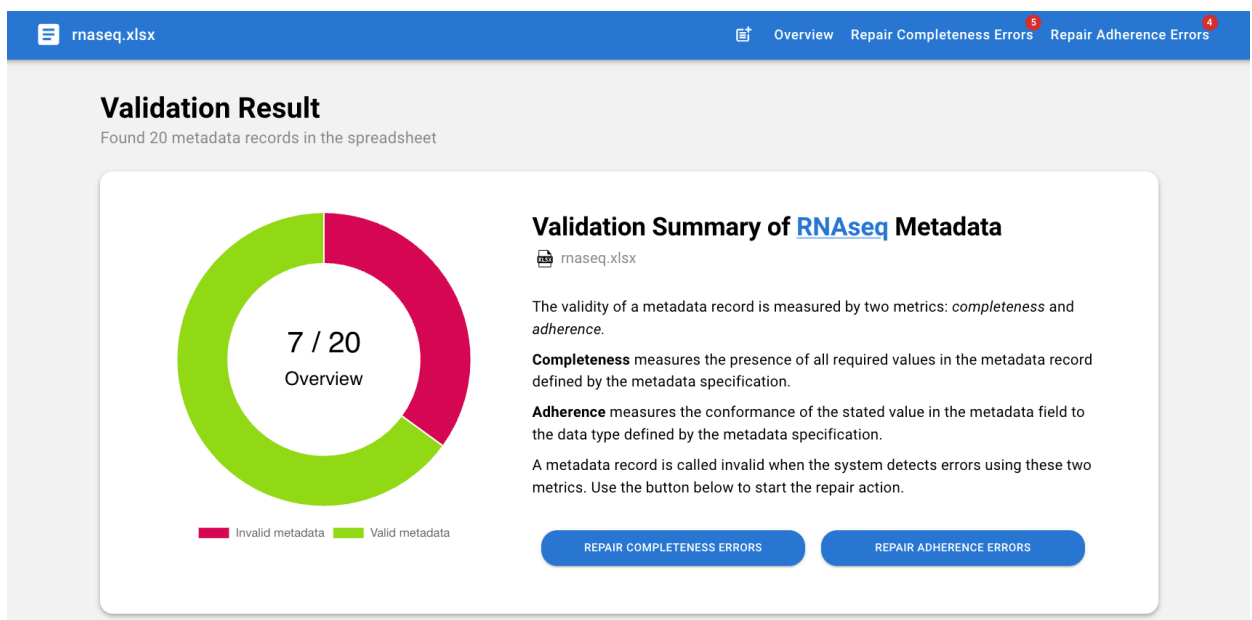

**Figure 7:** The CEDAR Metadata Validator to ensure the correctness of spreadsheet data. The Validator takes as input a metadata spreadsheet such as the one in Figure 6 and uses the corresponding CEDAR template to analyze the metadata to identify (1) missing values and (2) values that deviate from the standard. When there are errors in the metadata, the system can suggest possible fixes for the user to consider. The approach is used in HuBMAP to ensure that spreadsheet-based metadata remain adherent to the consortium's standards.

## Discussion

For the past decade, our group has been developing computer systems that assist with the management of research data. From the Web forms in the integrated CEDAR Workbench, to the distributed approach supported by the CEDAR Embeddable Editor, to the spreadsheets and validator system that support data stewardship for consortia such as HuBMAP, we have engineered widely used systems that make it easy to enforce the use of metadata standards when annotating datasets. Investigators feel enormous pressure from sponsors, from publishers, and from their peers in the scientific community to ensure that their data are FAIR—at a time when there is widespread confusion over what it is that really makes data FAIR in the first place. All our CEDAR-derived systems have been developed with the philosophy that, ultimately, the only component of FAIR data over which scientists have direct control is whether the metadata that annotate datasets adhere to standards endorsed by the relevant scientific community (Musen et al., 2022). We consequently see a great advantage in making metadata standards for scientific datasets machine actionable, and in making it possible not only for computers to reason about the standards and their application, but also to drive a range of metadata-management systems from a shared representation that is itself standardized.

In our work, we posit that scientific communities will want to develop shared models for metadata standards, representing those models as templates that enumerate the attributes of experiments in their discipline—attributes that capture information about the investigators, the experimental conditions, the subjects of the experiment, and the collected data. We see this approach being adopted widely in biology, and we anticipate that the notion of discipline-specific reporting guidelines for scientific datasets will take hold throughout much of science, particularly as the pressure for FAIR data ratchets upwards.

"Good old-fashioned AI," as used by CEDAR, provides a highly effective means for dealing with the increasing requirements for FAIR data. There is nothing in the FAIR guiding principles, however, that dictates the use of semantic technology. The FAIR principles mandate community-endorsed metadata standards, but they do not indicate how metadata should be structured or how they should be shared. The FAIR principles indicate that, when possible, the values of metadata fields should be supplied by terms from standard ontologies (specifically, from ontologies that are themselves FAIR), and yet there is no discussion of how ontology terms and value sets should be represented within a metadata specification. It is hard to imagine implementing solutions to ensure FAIR data without using traditional knowledge-representation techniques, however. In CEDAR, we adopted a lightweight approach to knowledge representation that we knew would be widely accepted by the user community that we were targeting.

CEDAR models metadata templates in JSON Schema, incorporating a pragmatic but principled approach to metadata representation. The CEDAR template model, consisting of metadata fields, composite entities, and overall templates (see Figure 2), respects the real-world needs of scientific communities while delivering on the promises of semantic rigor, reusability, and machine-actionability. The model supports a simple framework that represents metadata as lists of attribute–value pairs, where some groups of attributes may be reused across templates and where those groups of attributes may themselves include reusable groups of other attributes, recursively. By treating metadata standards not just as documents but as knowledge structures, CEDAR provides a solid foundation for building metadata-aware systems that are FAIR by design.

A CEDAR metadata template is a simple kind of knowledge base. Like the knowledge bases of the expert systems that the AI community pioneered decades ago (Hayes-Roth, et al., 1983), CEDAR templates capture in symbolic form the specialized beliefs of some community of practice. The knowledge bases of expert systems encoded beliefs about how professional tasks should be performed (e.g., how to diagnose and treat infection, how to configure the components of a computer system, where to drill for oil).  They encoded propositions about a task to be performed such that, when some process was applied to those propositions, the system would demonstrate behaviors that observers would judge to result in solutions on par with those that human experts might offer.  The process of constructing a knowledge base was termed *knowledge engineering* or *knowledge acquisition*, and early workers in the field often claimed that such knowledge bases allow computers to solve complex tasks in a manner akin to the way human experts do (Musen, 1993).  We now recognize that expert-system knowledge bases at best are crude models of how a person might solve a task (Clancey, 1989), but the development of such models is nevertheless an important achievement, and many expert systems did—and continue to do—remarkably important things.

CEDAR templates, taken by themselves, do not help to solve tasks.  They contain no knowledge of how to address domain-specific problems.  Still, they are knowledge bases—representations that capture the preferences of a community of practice regarding how scientific data should be described to enhance data FAIRness.

The groups of scientists who create and use CEDAR templates typically have distinct opinions regarding how their datasets should be annotated with metadata.  They understand what a third party needs to know in order to make sense of a dataset.  They often have strong beliefs about what needs to be said about the experimental conditions and the subjects of the study for someone to appreciate what was done and to grasp the context in which the data were collected.  Most important, they have a basic intuition for how such descriptions can be rendered as the attribute–value pairs that constitute modern experimental metadata.  The beliefs of the scientific community regarding how to constitute discipline-specific metadata translate directly into CEDAR templates, which, like classical expert-system knowledge bases, store information about how members of a professional community think about their work and about how they prefer to describe the results of their skilled activities.

CEDAR metadata templates are declarative representations, and, like more traditional knowledge bases, they can be reused for a variety of purposes.  As we have discussed, they can be used to render Web forms, to generate spreadsheets, and to inform systems that validate metadata entries.  In other work, we have also shown that CEDAR templates can enhance the display of metadata (Martínez-Romano, et al., 2025) and that they can markedly improve the performance of large language models asked to convert legacy metadata to a form that is closer to a given community standard (Sundaram et al., 2025).  CEDAR templates thus offer a canonical form in which a community can store and reuse its preferences for metadata structure and content, making those preferences inspectable by humans and actionable by machines; they are very much like the knowledge bases that we created in the 1980s.

Developers create CEDAR templates very much as we used to build knowledge bases for expert systems: Subject-matter experts work with colleagues who understand knowledge representation to iron out the details of what needs to be encoded and collaboratively build the necessary data structure. The GO FAIR Foundation (2025), in particular, has worked to develop stock methods for creating CEDAR templates through what they term *Metadata for Machines Workshops*. These meetings, usually held over several days, comprise intensive knowledge-engineering sessions that lead groups of scientists through the development of reporting guidelines for different classes of experiments, and that translate those guidelines into functioning CEDAR metadata templates. In these workshops, the evolving CEDAR templates serves (1) as a blackboard for capturing the participants' ongoing thoughts about what the metadata might describe and, at the end of the activity, (2) as the consensus standard for emerging from the group's deliberations. As is the case with all knowledge-engineering activities, the Metadata for Machines Workshops can be arduous and sometimes contentious, but the GO FAIR workers have developed a process for managing these sessions that is streamlined as much as possible.

Consortia such as HuBMAP adopt a more open-ended approach. HuBMAP has a Data Coordination Working Group (DCWG) that comprises both subject-matter expert and knowledge-representation specialists. The group works collaboratively to articulate reporting guidelines for different classes of biological assays, to render them as CEDAR templates, and to evaluate the use of the templates in practice. In the past three years, the DCWG has created some three dozen templates in this manner. Although the process of developing these community standards was often more iterative and sometimes more contentious than the DCWG would have liked, there really was no alternative if the goal was to ensure that the consortium's thousands of datasets would be FAIR.

Knowledge engineering has always had a bad reputation. It can be extremely labor-intensive. It can be difficult to model things in a manner to which all parties agree. It can be difficult for the process to scale. The advent of modern machine-learning methods in the past decade has caused the AI community to focus on data-driven approaches to the construction of intelligent systems, and traditional knowledge engineering seems to be becoming a lost art. Still, when the goal is to capture information that has never previously been expressed in a "learnable" form, manual knowledge engineering that involves intensive interaction with subject-matter experts becomes essential. It is impossible to learn standards for metadata from extant datasets that have been developed without consideration for the need for metadata standards in the first place. In such situations, there is no alternative than for subject-matter experts and domain modelers to sort things out.

The creation of knowledge bases of community-based standards for data annotation is bringing great benefits to different groups of scientists. When a consortium such as HuBMAP can render its standards as CEDAR templates, those templates become a detailed, examinable reference for those standards (HuBMAP Consortium, 2025). The templates help to ensure that metadata can be standardized from the moment that they are created, and that the datasets in the HuBMAP data repository are guaranteed to be FAIR. Such an ecosystem for data management encourages both the development and the adoption of community-based metadata standards, and, with the right infrastructure, it makes it almost trivial for scientists to adhere to the increasing mandates for FAIR data. If a research community is serious about open science, data sharing, and data reuse, it is not clear that there is any alternative to the approach offered by CEDAR. The good news is that we have several decades of experience in knowledge engineering that can inform the development of metadata knowledge bases. The major challenge is that most areas of science lack discipline-specific standards for descriptive metadata, and a continuing change in culture is needed to make the development of such standards a priority.

Scientific culture does change, however. No one questions the importance of preprints, of open access, or of bibliometrics, which are all relatively recent advances. No one can imagine a world in which investigators did not search the scientific literature on their own. FAIR data will become ubiquitous with the creation of a seamless infrastructure that makes it easy for investigators to develop and apply metadata standards in a transparent manner. And, most surprisingly, those advances will be rooted in AI technology that many of us thought was on the way out.

## Conclusion

Sponsors, publishers, and the scientific community are increasingly demanding that research data be findable, accessible, interoperable, and reusable (FAIR). FAIR data are possible only if research results are described using comprehensive, standardized, discipline-specific metadata that make it possible to search for datasets systematically, to integrate datasets from different sources, and to know what the investigators actually did. CEDAR is technology that can represent a scientific community's preferences for how to standardize metadata in the form of reusable templates. Experience over the past decade shows how CEDAR templates can drive a range of data-management applications. There is an urgent need to develop an infrastructure for science that makes such metadata templates accessible to investigators in a transparent manner, and that stimulates research communities both to create comprehensive metadata standards and to represent those standards as declarative knowledge bases.

## Acknowledgments

This work has been supported in part by grants U54 AI117925, R01 LM013498, and U24 GM143402, and by awards OT2 OD033759 and OT2 DB000009 from the National Institutes of Health, by grants 1937698 and 2134956 from the National Science Foundation, and by a grant from the Templeton World Charity Foundation. We are grateful to Mete Akdoğan and Attila Egyedi for their outstanding contributions to the CEDAR project.